\documentclass[aps, prl, twocolumn, reprint, showkeys, superscriptaddress, longbibliography]{revtex4-2}

\usepackage{amsmath}
\usepackage{amssymb}
\usepackage[T1]{fontenc}
\usepackage[utf8]{inputenc}
\usepackage{graphicx}
\usepackage[dvipsnames]{xcolor}
\usepackage[caption=false]{subfig}
\usepackage[colorlinks=true, linkcolor=blue, citecolor=blue, filecolor=blue, urlcolor=blue, pdfproducer={.}, pdfcreator={.}]{hyperref}

\usepackage{orcidlink}

\usepackage{siunitx}
\DeclareSIUnit\ML{ML}
\DeclareSIUnit\angstrom{\text{\AA}}
\DeclareSIUnit\amunit{u} 
\DeclareSIUnit\mbar{mbar}

\newcommand{\Tc}{T_\mathrm{c}}
\newcommand{\kB}{k_\mathrm{B}}
\newcommand{\Eb}{E_\mathrm{b}}
\newcommand{\DE}{\Delta E}
\newcommand{\CV}{C_\mathrm{V}}
\newcommand{\TD}{\Theta_\mathrm{D}}
\newcommand{\Svib}{S_\mathrm{vib}}
\newcommand{\Sconf}{S_\mathrm{conf}}
\newcommand{\Uth}{U_\mathrm{th}}
\newcommand{\eightbytwo}{(8{\times}2)}
\newcommand{\fourbyone}{(4{\times}1)}

\begin{document}



\title{Zero-Temperature Energetics of a First-Order Surface Phase Transition from Diffraction}

\author{Mohammad Tajik \orcidlink{0000-0003-3093-9142}}
\email{meta.tajik6889@gmail.com}
\affiliation{Faculty of Physics, University of Duisburg-Essen, 
47057 Duisburg, Germany}

\author{Christian Brand \orcidlink{0000-0002-2984-4456}}
\affiliation{Faculty of Physics, University of Duisburg-Essen, 
47057 Duisburg, Germany}

\author{Stefan Wippermann \orcidlink{to be asked}} \affiliation{Department of Physics, Philipps University Marburg, 35032 Marburg, Germany}

\author{Wolf Gero Schmidt \orcidlink{to be asked}} \affiliation{Department of Physics, Paderborn University, 33098 Paderborn, Germany}

\author{Michael Horn-von Hoegen \orcidlink{0000-0003-0324-3457}}
\email{mhvh@uni-due.de}
\affiliation{Faculty of Physics, University of Duisburg-Essen, 
47057 Duisburg, Germany}
\affiliation{Center for Nanointegration Duisburg-Essen (CENIDE), 
47057 Duisburg, Germany}

\begin{abstract}
First-order phase transitions are governed by the balance between zero-temperature energy, thermal vibrational energy, and entropy, yet these contributions are rarely accessible experimentally.
Here, we establish a diffraction-based route to determine them for the prototypical Si(111)--In atomic-wire system.
From the temperature-dependent Debye--Waller attenuation of reflection high-energy electron diffraction (RHEED) intensities we extract distinct Debye temperatures for the low-temperature $\eightbytwo$ ground state and the high-temperature $\fourbyone$ phase.
The softer vibrational spectrum of the high-temperature phase yields the entropy gain that stabilizes this phase above the critical temperature $\Tc$.
Evaluating the thermodynamic balance between internal energy and entropy at $\Tc = \SI{125}{\K}$ yields a zero-temperature energy difference of $\DE_0 = \SI{138}{\meV}/\eightbytwo$, including the difference in vibrational zero-point energies of the two phases, thus providing direct access to hidden thermodynamic parameters of a first-order surface phase transition. 
\end{abstract}

\maketitle

\par Discontinuous phase transitions (PT) of first order are omnipresent in condensed matter.
Examples include solid-to-liquid PTs like the melting of ice into water \cite{thomson1849theoretical}, the metal-to-insulator PT of V$_2$O$_3$ \cite{mcwhan1973metal, mcwhan1969mott}, the ferroelectric structural PT in BaTiO$_3$ \cite{devonshire1949xcvi}, the antiferromagnetic-to-ferromagnetic PT in FeRh \cite{fallot1938alliages}, the magnetostructural PT in MnAs \cite{de1963magnetic}, and amorphous-to-crystalline PT in phase change materials \cite{wuttig2007phase}.
Displacive structural PTs often occur due to breaking/formation of specific atomic bonds or small changes in the bonding angle between atoms.
Electronic transitions also belong to the class of first-order PT, including the magnetic-field-driven PT from the Meissner state to the normal state in type-I superconductors \cite{GinzburgLandau1950} and the correlation-driven Mott metal--insulator PT \cite{Mott1968}.

\par A first-order PT from a low temperature (LT) to a high temperature (HT) phase is marked by phase coexistence and a finite energy barrier $\Eb$ between the two phases at the critical temperature $\Tc$.
The barrier $\Eb$ gives rise to phenomena like metastability, supercooling, and a hysteresis of the order parameter upon temperature cycling.
The PT is thermodynamically discontinuous: the first derivative of the Landau free energy exhibits discontinuities, such as abrupt changes in volume or entropy at $\Tc$ and gives rise to latent heat.

\par All these characteristic properties of a first-order PT can be described in the framework of the Landau theory \cite{landau2013course}.
Specifically, for a displacive PT the Landau potential $\phi(x)$ can be expanded in even powers of the displacive coordinate $x$ as
\begin{equation}
\phi(x) = \kappa_2 x^2 - \kappa_4 x^4 + \kappa_6 x^6 ,
\label{eq:potential-energy}
\end{equation}
with all coefficients $\kappa_i > 0$.
The essential feature is the negative quartic term together with the positive sixth-order term, which stabilizes the free energy for large displacements of $x$ \cite{fultz2020phase}. 

\par
\begin{figure}[b]
  \centering
  \includegraphics[width=\columnwidth]{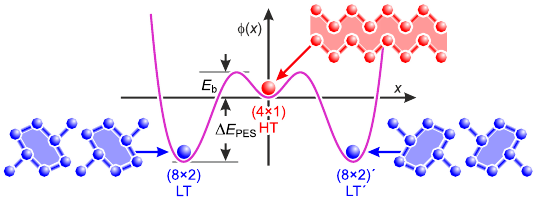}
  \caption{
  Sketch of Landau potential energy $\phi(x)$ as a function of displacement $x$ at $T = \SI{0}{K}$, representing the energetically favored $\eightbytwo$ LT ground states, the higher-energy $\fourbyone$ HT configuration, the static PES difference $\DE_\mathrm{PES}$, and the transition barrier $\Eb$.
  Here, the potential energy surface for the indium atomic wire system has been reproduced which was calculated by DFT \cite{wall2012atomistic}.
  The thermodynamic zero-temperature energy difference determined below is $\DE_0 = \DE_\mathrm{PES} + \Delta U_\mathrm{ZP}$ and therefore also contains the difference in vibrational zero-point energies of the two phases.
  }
  \label{Landau}
\end{figure}

 A typical Landau potential is sketched in Fig.~\ref{Landau}.
It exhibits two deep minima $ \phi(\pm x_\mathrm{LT})$ for the LT state at large displacements $\pm x_\mathrm{LT}$ and an additional minimum for the HT state, $\phi(x_\mathrm{HT}) = 0$, at the high-symmetry position $x_\mathrm{HT} = 0$.
These (meta)-stable phases are separated by energy barriers $\Eb$.
The static potential-energy difference between the two states is $\DE_\mathrm{PES} = \phi_\mathrm{HT} - \phi_\mathrm{LT}$. 

\par At finite temperatures, vibrational zero-point motion, thermally excited phonons, and entropy contribute to the thermodynamic stability.
For each phase $i = \mathrm{LT}, \mathrm{HT}$, the free energy at the corresponding minimum is therefore written as
\begin{equation}
F_i(T) = E_{0,i} + U_{\mathrm{th},i}(T) - T S_i(T)\,,
\label{eq:free-energy}
\end{equation}
where $E_{0,i} = \phi_i + U_{\mathrm{ZP},i}$ is the zero-temperature energy including the vibrational zero-point contribution, and $U_{\mathrm{th},i}(T) = \int_0^T C_{\mathrm{V},i}(T')\,\mathrm{d}T'$ denotes the thermal vibrational energy above the zero-point level. 
The corresponding zero-temperature energy difference is \linebreak $\DE_0 = E_{0,\mathrm{HT}} - E_{0,\mathrm{LT}}$.
Microscopically, the two coexisting phases of a solid are separated by $\Eb$ and only at $\Tc$ their Landau free energies are identical \cite{landau2013course, moulson2003electroceramics, porter2009phase, fultz2020phase}.

\par However, the thermal vibrational and entropic contributions are usually not known for all phases.
For a PT under equilibrium conditions, it is sufficient to consider the free energies at the relevant minima.
The experimental and/or theoretical determination of decisive parameters such as $\DE_0$ or $\Eb$ is, however, far from trivial.

\par Only isolated calorimetric studies have previously addressed related energy differences. The energy difference between reconstructed Pt(100) surface phases has been measured calorimetrically \cite{Yeo1995}, while the zero-temperature energy difference associated with the AFM--FM transition in FeRh has been inferred using a compositionally modified ferromagnetic reference sample \cite{Cooke2012}. However, neither approach determines the zero-temperature energy difference from phase-resolved vibrational thermodynamics obtained by diffraction. Here, we demonstrate that this quantity can be determined from phase-specific Debye--Waller measurements and the resulting vibrational thermodynamics.
Neither approach determines the zero-temperature energy difference from phase-resolved vibrational thermodynamics obtained by diffraction.
Here, we show that this quantity can be obtained from phase-resolved Debye--Waller measurements and the resulting vibrational thermodynamics.

\par For the prototypical first-order PT of the In atomic-wire system on Si(111), we disentangle the contributions of zero-temperature energy, thermal vibrational energy, and entropy to the free energy.
We first exploit the Debye--Waller attenuation of the diffraction intensities to determine effective phase-specific Debye temperatures.
These experimentally determined vibrational energy scales are then used within a two-dimensional Debye thermodynamic model to obtain the heat capacity $\CV$, thermal vibrational energy $\Uth$, and vibrational entropy $\Svib$.
This approach yields $\DE_0$ and thereby provides access to thermodynamic properties of the first-order PT that are otherwise difficult to determine experimentally.

\par The structural PT for this monolayer adsorbate system is characterized by a reversible change of the surface structure from LT $\eightbytwo$ to HT $\fourbyone$ reconstruction.
Above $\Tc^+ = \SI{130}{\K}$ the In atoms arrange themselves in zigzag chains with a $\fourbyone$ periodicity as is sketched in Fig.~\ref{Landau}.
Upon cooling below $\Tc^- = \SI{120}{\K}$, the surface undergoes a symmetry-breaking structural PT to the $\eightbytwo$ periodicity.
The zigzag chains of In atoms break up and instead they arrange in distorted hexagons \cite{yeom1999instability, yeom2002nature} as is sketched in Fig.~\ref{Landau}.
This structural PT is accompanied by the formation of a charge density wave (CDW) \cite{lee2004defect, kim2016origin, lee2005real, wall2012atomistic, frigge2017optically}.
The difference in the critical temperature during heating $\Tc^+$ and cooling $\Tc^-$ causes a thermal hysteresis $\Delta T = \Tc^+ - \Tc^- \simeq \SI{10}{\K}$.
The width of the hysteresis $\Delta T$ is independent of the cooling/heating rate $\mathrm{d}T / \mathrm{d}t$ \cite{klasing2014hysteresis, hafke2019condensation}.
Both findings are indicative of a first-order PT.
For this system, the PES as function of a generalized reaction coordinate $R_\mathrm{grc}$ has been determined by density functional theory with a static (neglecting zero-point vibrations) PES difference $\DE_{\mathrm{PES,DFT}} = \SI{96}{\meV}/\eightbytwo$ and $\Eb = \SI{40}{\meV}/\eightbytwo$ \cite{wall2012atomistic}. 

\par In addition to the structural transformation, the Si(111)--In system has been reported to undergo a metal-to-insulator transition upon cooling below $\Tc$, accompanied by the opening of an electronic band gap $E_\mathrm{g}$ \cite{yeom2002nature, ahn2004mechanism, tanikawa2004surface, im2023topological}.
This gap opening is commonly discussed in terms of a Peierls-like instability, in which the distortion of the LT phase doubles the periodicity along the two in-plane surface directions, transforming the $\fourbyone$ reconstruction into the $\eightbytwo$ phase \cite{wall2012atomistic}.
Conversely, the HT phase above $\Tc$ is stabilized by its larger vibrational entropy \cite{wippermann2010entropy}.
However, reported experimental and theoretical values of $E_\mathrm{g}$ span a wide range, from approximately \qtyrange{100}{400}{\meV} \cite{yeom1999instability, nicholson2019excited, uetake2012anisotropic, im2023topological}, indicating that $E_\mathrm{g}$ alone does not provide a unique thermodynamic measure of the PT.

\par The In atomic wire system is prepared by self-assembly under ultra-high vacuum conditions of \linebreak $p = \qty[tight-spacing = true]{2e-10}{\mbar}$.
\textit{In-situ} deposition of a In monolayer ($\SI{1}{\ML} = \qty[tight-spacing = true]{7.83d14}{\mathrm{atoms}\per\cm^2}$) on a clean Si(111) surface at a sample temperature of \qtyrange{700}{750}{\K} creates the Si(111)--$\fourbyone$-In reconstruction \cite{yeom1999instability, yeom2002nature, ahn2004mechanism, kumpf2000low, sakamoto2000angle, park2004direct, guo2005intertwined}.
Heating and cooling of the Si sample were facilitated through a temperature-controlled liquid He-cooled cryostat (Janis, Lakeshore).
The change of surface structure during the PT was followed by surface sensitive electron diffraction.
We used high-resolution reflection high-energy electron diffraction (RHEED) \cite{hafke2019pulsed} to measure the intensity of $\fourbyone$ and $\eightbytwo$ spots during temperature cycling at a rate of $\SI{0.07}{\K/\s}$ in the range from \qtyrange{30}{210}{\K}.
An entire heating/cooling cycle was performed in only \SI{80}{\minute}, thus minimizing adsorption from residual gas which otherwise is detrimental for the PT as it broadens and shifts the PT \cite{lee2009roles, shibasaki2010phase}.

\begin{figure*}[t]
  \centering
  \includegraphics[width=\textwidth]{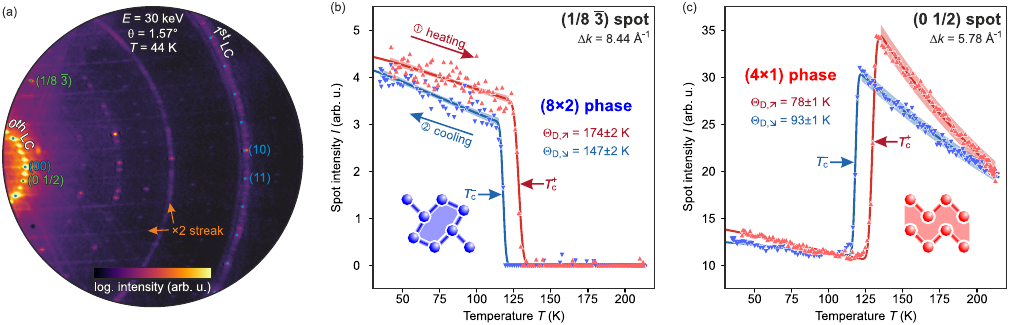}
  \caption{RHEED measurements of the In atomic-wires during the first-order structural PT.
  (a) Diffraction pattern of the LT $\eightbytwo$ phase showing diffraction spots and ${\times2}$ streaks among all three rotational domains of the In atomic wires on Si(111) at $T = \SI{44}{\K}$.
  Integer order spots are indicated by blue dots on the 0th and 1st Laue circle (LC), green dots indicate the positions of the $\left( \frac{1}{8}~\bar{3} \right)$ and $\left( 0~\frac{1}{2} \right)$ spot shown in (b,c).
  (b) Relative intensity of a representative LT $\eightbytwo$ spot measured during heating and cooling.
  The abrupt drop and recovery of the $\eightbytwo$ intensity indicate the melting and recrystallization of the LT phase, exhibiting a pronounced thermal hysteresis $\Delta T$ between heating and cooling.
  (c) Relative intensity of a representative HT $\fourbyone$ spot measured during the same thermal cycle.
  Increase and decrease of the $\fourbyone$ intensity shows the formation and suppression of the HT phase.
  Solid lines in (b,c) represent fits for the Debye temperatures $\TD^{\eightbytwo}$ and $\TD^{\fourbyone}$, while the shaded areas are the $5\sigma$ uncertainties of the fits.
  The raw data are binned to reduce the noise in the graph.
  Ball-and-rod models depict the characteristic distorted hexagons of the $\eightbytwo$ ground state and the characteristic zig-zag chains of the $\fourbyone$ state.}
  \label{Fig:RHEED}
\end{figure*}

\par Figure~\ref{Fig:RHEED}(a) shows the diffraction pattern of the LT $\eightbytwo$ phase taken at an electron energy of $E = \SI{30}{\keV}$ and at an angle of incidence of $\theta = \SI{1.57}{\degree}$.
The pattern consists of contributions from all three rotational domains of the In atomic wires on Si(111).
Fig.~\ref{Fig:RHEED}(b,c) show the spot intensities of two representative diffraction spots as a function of temperature $T$ upon heating (reddish triangles) and cooling (bluish triangles).
The intensity of the $\eightbytwo$ spot is shown in Fig.~\ref{Fig:RHEED}(b).
It vanishes upon heating at $\Tc^+ = \SI{130}{\K}$ and reappears upon cooling at $\Tc^- = \SI{120}{\K}$.
This behavior reflects the melting of the LT CDW ground state and the lifting of the Peierls distortion.
The hysteresis upon heating and cooling is indicative for a first-order PT.
For $T < \Tc$, the intensity slightly rises with decreasing $T$, which is explained by the Debye--Waller effect,
\begin{equation}
I / I_0 = \exp \left( - 1/3 \left\langle \textbf{u}^2 \right\rangle \Delta \textbf{k}^2 \right),
\label{eq:DW}
\end{equation}
with the mean squared displacement $\left\langle \textbf{u}^2 \right\rangle$ of the thermal motion of the atoms and the momentum transfer $\Delta \textbf{k}$ of the diffracted electrons.
Here, we analyze the $\left( \frac{1}{8}~\bar{3} \right)$ spot with $| \Delta \textbf{k} | = \SI{8.44}{\angstrom^{-1}}$. 
Within the conventional Debye--Waller approximation, the mean-squared displacement $\left\langle  \textbf{u}^2 \right\rangle$ is related to an effective surface Debye temperature $\TD$ according to
\begin{equation}
\left\langle \textbf{u}^2 \right\rangle = \frac{9 \hbar^2}{M \kB \TD} \left[ \frac{1}{4} + \frac{T}{\TD} \varphi\!\left( \frac{\TD}{T} \right) \right]\,,
\label{eq:MSD}
\end{equation}
with the atomic mass $M = \SI{114.8}{\amunit}$ for the In atoms, the Debye function of first order $\varphi(x) = \frac{1}{x} \int_0^x \xi / ( \mathrm{e}^\xi - 1 )\,\mathrm{d}\xi$, and including the zero-point contribution to the mean squared atomic displacement \cite{WillisPryor1975}.
We evaluated several diffraction spots and weighted the individual Debye temperatures with the spot intensity to reduce the influence of dynamic scattering effects.
Our analysis yields $\TD^{\eightbytwo} = \SI[separate-uncertainty-units=single, tight-spacing=true]{159 \pm 13}{\K}$.

\par The PT is modeled by a sinusoidal function with its inflection point at $\Tc^+$ and $\Tc^-$.
The systematic difference in intensity upon heating and cooling is explained by adsorption from the residual gas, which affects the long-range order at the surface and thus slightly reduces the spot intensity during the course of the experiment. 

\par Accordingly, the formation of the HT phase is followed through the intensity of the $\fourbyone$ spots as a function of temperature.
Fig.~\ref{Fig:RHEED}(c) shows a rise of intensity for the $\left( 0~\frac{1}{2} \right)$ spot with $| \Delta \textbf{k} | = \SI{5.78}{\angstrom^{-1}}$ during temperature cycling.
Upon heating, the intensity rises at $\Tc^+ = \SI{130}{\K}$ and drops upon cooling at $\Tc^- = \SI{120}{\K}$, reflecting the formation of the HT $\fourbyone$ phase.
With increasing temperature, the intensity decreases further, again explained by the Debye--Waller effect.
Applying the same analysis as for the $\eightbytwo$ spots, we obtain $\TD^{\fourbyone} = \SI[separate-uncertainty-units=single, tight-spacing=true]{86 \pm 11}{\K}$ for the HT $\fourbyone$ phase after averaging.
Both Debye temperatures are in close agreement with previous measurements \cite{RHEPD-InSi}.

\begin{figure*}[t]
  \centering
  \includegraphics[width=\textwidth]{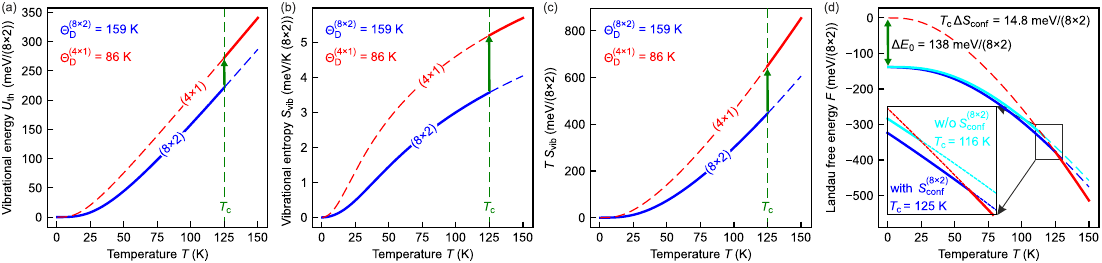}
  \caption{
  Thermodynamic states plotted as a function of temperature for the LT $\eightbytwo$ phase (blue solid and dashed lines) and for the HT $\fourbyone$ phase (red solid and dashed lines) per $\eightbytwo$ unit cell.
  Debye temperatures of $\TD^{\eightbytwo} = \SI[separate-uncertainty-units=single, tight-spacing=true]{159 \pm 13}{\K}$ and $\TD^{\fourbyone} = \SI[separate-uncertainty-units=single, tight-spacing=true]{86 \pm 11}{\K}$ were applied for the $\eightbytwo$ and $\fourbyone$ phases, respectively.
  The green dashed line and arrow indicate the critical temperature $\Tc = \SI{125}{\K}$ and the phase differences in $\Uth$, $\Svib$, and $T \Svib$ at $\Tc$.
  (a) Thermal vibrational energy $\Uth$, measured relative to the respective zero-point level.
  (b) Vibrational entropy $\Svib$. 
  (c) Entropic contribution $T \Svib$ to the free energy. 
  (d) Free energies $F$ of the LT $\eightbytwo$ and HT $\fourbyone$ phases.
  The zero-temperature energy offset is $\DE_0 = \SI{138}{\meV/\eightbytwo}$ when the upper bound $\Sconf^\mathrm{max} = \kB \ln 4$ is used.
  The cyan curve shows the limiting case $\Sconf = 0$, illustrating the maximal effect of the configurational contribution on the free-energy crossing.
  The inset shows in enlarged scale the crossovers of the Landau free energy around $\Tc$ together with the contribution of configurational entropy.
  }
  \label{enthalpy-entropy}
\end{figure*}

\par The pronounced difference in $\TD^{\eightbytwo}$ compared to $\TD^{\fourbyone}$ confirms the significant blue shift of the vibrational spectrum for the LT phase when compared with the HT phase \cite{fleischer2007surface, wippermann2010entropy, PhysRevB.94.075417}.
This difference is a key requirement for the existence of a PT: although the LT phase is favored by its lower zero-temperature energy, as \linebreak $\DE_0 > 0$, the HT phase becomes stable at $T > \Tc$ through its larger vibrational entropy, $\Svib^{\fourbyone} > \Svib^{\eightbytwo}$, which compensates for the higher zero-temperature energy of the HT phase in the Landau free energy balance [Eq.~\eqref{eq:free-energy}]. 
To quantify these terms, we derive the vibrational entropy $\Svib(T) = \int_0^T \CV(T') / T'\,\mathrm{d}T'$ and the thermal vibrational energy $\Uth(T) = \int_0^T \CV(T')\,\mathrm{d}T'$ from the heat capacities of the two phases.
For a solid surface, where thermal expansion effects are small at temperatures considered here, we approximate the experimentally relevant heat capacity by $C_\mathrm{p} \simeq \CV$.
Since $\CV(T)$ changes strongly for $T < \TD$, its full temperature dependence is retained and modeled within a two-dimensional Debye model.
In this model, the low-temperature heat capacity follows $C_\mathrm{p} \propto T^2$, while in the high-temperature limit it approaches $2N \kB$ instead of the three-dimensional Dulong--Petit limit of $3N \kB$ \cite{huang2008theoretical}:
\begin{equation}
\CV(T) = 4 N \kB \left( \frac{T}{\TD} \right)^2 \int_0^{\TD / T} \frac{x^3 \mathrm{e}^x}{\left( \mathrm{e}^x - 1 \right)^2}\,\mathrm{d}x\,.
\label{eq:cv-2d-debye}
\end{equation}

\par Figure~\ref{enthalpy-entropy} depicts the thermodynamics of the In atomic wire system.
The thermal vibrational energy $\Uth(T)$ is shown in Fig.~\ref{enthalpy-entropy}(a) for the LT $\eightbytwo$ and HT $\fourbyone$ phase as blue and red lines, respectively.
Due to the enhanced stiffness of the LT $\eightbytwo$ phase we find \linebreak $\Uth^{\eightbytwo}(T) < \Uth^{\fourbyone}(T)$ for all temperatures.
At $\Tc = \SI{125}{\K}$, the thermal vibrational energies of the two phases differ by $\Delta \Uth(\Tc) = \SI{51}{\meV/\eightbytwo}$ (vertical green arrow).
The vibrational entropy $\Svib(T)$ is shown in Fig.~\ref{enthalpy-entropy}(b) and differs between the two phases by $\Delta \Svib(\Tc) = \SI{1.63}{\meV/\K/\eightbytwo}$ at $\Tc$ (vertical green arrow), as expected for a first-order PT.
The temperature-scaled vibrational entropy, $T \Svib$ contribution to the free energy is shown in Fig.~\ref{enthalpy-entropy}(c).  

\par So far, only the vibrational contribution $\Svib(T)$ has been considered.
The symmetry-broken LT phase, however, also carries a configurational entropy $\Sconf^{\eightbytwo}$.
While only one configuration exists for the high-symmetry HT state, four symmetry-related configurations are possible for the symmetry-broken LT state.
These four geometric configurations are built by lateral shifts by one lattice unit along the wires and by the two mirror-related orientations of the distorted hexagons in adjacent rows. 
Experimental evidence for these four configurations is provided by the extended streaks observed at the half-order positions of the $\eightbytwo$ reconstruction, rather than by discrete diffraction spots \cite{klasing2014hysteresis}.
The streaks indicate incomplete long-range correlation of the positions and orientations of the distorted hexagons across the wire array, although short-range correlations may persist.
Therefore, the four symmetry-related configurations provide an upper bound for the configurational entropy $\Sconf^\mathrm{max} = \kB \ln 4$, that lowers the Landau free energy.
The actual configurational entropy is reduced by spatial correlations, in particular by the high formation energy of domain walls within individual wires \cite{razzaq2020dualnaturelocalizedphase}, whereas lower-energy phase slips and boundaries between neighboring wires may provide a residual contribution.
At $\Tc$, this upper bound corresponds to $\Tc \Sconf^\mathrm{max} = \SI{4.8}{\meV/\eightbytwo}$, which is sufficiently large to be included in the thermodynamic balance.
Neglecting $\Sconf$ would cause a shift of $\Tc$ by \SI{-9}{\K} (see cyan dashed curve Fig.~\ref{enthalpy-entropy}(d)); therefore, $\Sconf$ may contribute to the stabilization of the LT $\eightbytwo$ phase.

\par We now have all contributions required to evaluate the free energy balance shown in Fig.~\ref{enthalpy-entropy}(d) as blue and red curves for the LT $\eightbytwo$ and HT $\fourbyone$ phases, respectively.
The large entropic contribution $T \Delta \Svib$ causes the downward slope for both phases, which are separated for $T = \SI{0}{\K}$ by the zero-temperature energy difference $\DE_0$, including the difference in vibrational zero-point energies, which stabilizes the LT phase.
With increasing temperature, the larger entropic contribution of the HT $\fourbyone$ phase compensates the higher thermal vibrational energy and the zero-temperature energy difference $\DE_0$ at $\Tc$.
Using $\Tc \simeq ( \Tc^+ + \Tc^- ) / 2 = \SI{125}{\K}$ and the upper bound $\Sconf^\mathrm{max} = \kB \ln 4$, we derive $\DE_0 = \SI{138}{\meV/\eightbytwo}$.
Static DFT calculations yield a PES difference of \linebreak $\DE_\mathrm{PES,DFT} = \SI{96}{\meV/\eightbytwo}$ \cite{wall2012atomistic}; a direct quantitative comparison requires inclusion of the phase-dependent vibrational zero-point energies in the theoretical value.

\par In conclusion, we have established a thermodynamic route to determine the zero-temperature energy difference $\DE_0$ between competing phases of a first-order surface PT.
For the Si(111)--In atomic-wire system, phase-specific Debye temperatures obtained from the Debye--Waller attenuation provide access to the thermal vibrational energy and entropy of the LT $\eightbytwo$ and HT $\fourbyone$ phases.
Combining these contributions with the configurational entropy of the LT phase and the condition of equal free energies at $\Tc = \SI{125}{\K}$ yields \linebreak $\DE_0 = \SI{138}{\meV/\eightbytwo}$, including the difference in vibrational zero-point energies of the two phases.
The experimental uncertainties of the Debye temperatures result in a range of $\qtyrange[range-units = single]{86}{193}{\meV/\eightbytwo}$. 

\par The larger vibrational entropy of the softer HT phase stabilizes the $\fourbyone$ structure at elevated temperatures despite its higher zero-temperature energy.
More generally, our results demonstrate that diffraction intensities, analyzed through the Debye--Waller effect and an appropriate thermodynamic model, can provide access to thermodynamic quantities beyond structural information and offer a route to investigate first-order phase transitions between entropy-stabilized competing phases.

\section*{author contributions}
M.T. conceived and implemented the research concept, performed the thermodynamic computations and the analysis.
C.B. and M.T. contributed to the Debye–Waller analysis.
S.W. contributed to the analysis of the configurational entropy.
M.T., W.-G.S., and M.H.-v.H. coordinated the work and manuscript preparation.
All authors contributed to writing and revising the manuscript.

\section*{Acknowledgements}
Experimental work and fruitful discussions with B. Krenzer and T. Frigge are gratefully acknowledged.
Funded by the Deutsche Forschungsgemeinschaft (DFG, German Research Foundation) -- Project-ID 278162697 -- SFB1242 ``Non-equilibrium dynamics of condensed matter in the time domain'' (project C03); Project-ID 575166306 -- SCHM1361/35.

\bibliography{bibliography_new}

\end{document}